\documentstyle[aps,epsf]{revtex}

\def\epsfig#1#2#3#4
         {
         \epsfysize=#2 \vbox{ \hglue#3 \epsfbox[#4]{#1} }
         }
\def\epsfigrot#1#2#3#4
         {
         \epsfxsize=#2
         \setbox\rotbox=\hbox to #2{\epsfbox[#4]{#1}}
         \vbox{\hglue#3 \rotl\rotbox}
         }
\newbox\rotbox

\begin{document}
\title{Tunneling in quantum wires I:  Exact solution of the spin isotropic 
case.}
\author{F. Lesage$^{1,2}$, H. Saleur$^{1,2}$, P. Simonetti$^{1,2}$.}
\address{${}^1$ Institute for theoretical physics, University of
California, Santa-Barbara, CA 93106-4030.}
\address{${}^2$ On leave from: 
Department of Physics, University of Southern California,
Los-Angeles, CA 90089-0484.}
\date{\today}
\maketitle

\begin{abstract}
We show that the  problem of  impurity tunneling in a Luttinger
liquid of electrons with spin is solvable  in the spin isotropic 
case ($g_\sigma=2$, $g_\rho$ arbitrary).  The resulting integrable model
is similar  to a two channel anisotropic Kondo model,
but  with the impurity spin
in a ``cyclic representation'' of the quantum algebra $su(2)_q$ 
 associated with
the anisotropy.
Using  exact,  non perturbative  techniques
we study the RG flow, and compute the DC conductance.
As expected from the analysis of Kane and Fisher we find that
the IR fixed point corresponds to two separate leads. 
We also prove  an exact duality between the
UV and IR expansions of the current at vanishing temperature.

\end{abstract}

\section{Introduction.}

Quantum impurity problems  have attracted
constant attention recently. The reason
is, that the underlying physics is highly non trivial,
that the models are manageable technically despite the
presence of very strong interactions, and that there
are many practical/experimental applications. The latter   include
the Kondo effect, quantum dots, dissipative quantum mechanics,
impurity tunneling in fractional quantum Hall devices and in one
dimensional quantum wires.

The physical variables  are usually the temperature,
the external field,  the bulk interactions, the nature of the
impurity
(which can either be a charge or spin impurity), and the way it is
coupled to the bulk degrees of freedom.
A standard approach to computing physical quantities in these
models would be to use perturbation theory in the bulk-impurity
coupling for example.  This perturbative expansion
is most of the time done around the UV fixed point,
and  usually  leads to limited results.
Since all the problems of interest are essentially one dimensional
(or can be reformulated as such),  more powerful methods can however
be used.
For instance, in recent years, a conformal field
theory description has allowed a much better understanding of
the fixed points and their vicinity  \cite{afflud}.

Another method, which  is a priori less general, is  the use of
quantum integrability, pioneered in the  study of the Kondo
problem \cite{wieg}. Integrability is a powerful, non perturbative
tool, which  is not limited
to the vicinity of the fixed points but allows a complete
description of the properties all the way along the
renormalisation group flow, from the UV to the IR fixed points.
Technical difficulties have for a long time  restricted the use of
integrability to the computation of thermodynamic, equilibrium
properties.
However, recent progress, in particular in the understanding of form
factors \cite{S} and
integrable description of conformal field theories \cite{FSZ,BLZ},
have  also made possible the computation
of dynamical, out of equilibrium  properties for many quantum
impurity problems \cite{flsbig,LSS}.

A still open challenge in that field is the problem of  a single
(charge) impurity
in a  one dimensional quantum wire, where the electrons are described
by a Luttinger liquid with charge and spin degrees of freedom.
This problem was studied using the renormalisation group and
perturbation theory by Kane and Fisher \cite{kanefish}
and Furusaki and Nagaosa \cite{furusaki}.  There, a
complete picture of the flows generated by the RG was
obtained in terms of the ``$g$" factors, $g_\rho$ and
$g_\sigma$, describing the interaction of  bulk charge and spin
degrees of freedom. The existence of tentalizing new fixed points,
which would be partly
transmitting and partly reflecting, was in particular conjectured.
Steps   were taken to identify these fixed points using conformal
field theory \cite{AW}, but the problem remains open.

In this paper,  we are addressing this impurity problem from the
point of view of integrability.  The case of an impurity in a
spinless electron gas could
be mapped on the boundary sine-gordon model \cite{flsbig}, which is a
completely integrable quantum field theory \cite{GZ}, formally 
analogous to the 
anisotropic Kondo problem, but with the impurity spin in a cyclic
representation of the quantum algebra $su(2)_q$ \cite{FLS}. 
Here, we show that the case of
electrons with spin  maps, when  $g_\sigma=2$,
onto  a model analogous to the anisotropic two channel Kondo model, 
again with the impurity spin chosen in a cyclic representation
of the quantum algebra $su(2)_q$. This allows us to compute the relevant 
physical properties.

\section{The model, integrability, and the $S$ matrix.}

In order to fix notations, let us repeat
some of  the basic definitions in \cite{kanefish}.
In the one dimensional Luttinger liquid, bosonisation of the
fermionic operators is accomplished via
\begin{equation}
\psi_\mu^\dagger \simeq \sum_{n \ odd} e^{i n[\sqrt{\pi} \Theta_\mu
+k_F x]} e^{i\sqrt{\pi} \phi_\mu(x)},
\end{equation}
with $\mu=\uparrow , \downarrow$.  The fields $\phi_\mu$ and
$\Theta_\mu$ have commutation relations
\begin{equation}
[\phi_\mu(x),\Theta_{\mu'}(x')]=i \delta_{\mu ,\mu'} \theta(x-x')
\end{equation}
from which we can devise two different representations of the
Luttinger liquid
(our conventions follow \cite{kanefish}).
Here, we will work in the so called $\Theta$-representation.
Changing basis to the charge and spin degrees of freedom
\begin{equation}
\Theta_\rho=\Theta_\uparrow+\Theta_\downarrow , \ \
\Theta_\sigma=\Theta_\uparrow-\Theta_\downarrow ,
\end{equation}
we obtain the action
\begin{eqnarray}
S&=&S_\rho+S_\sigma  \\
&=&\int dx dy \frac{1}{2g_\rho} [(\partial_x \Theta_\rho)^2+
(\partial_y \Theta_\rho)^2]+
\frac{1}{2g_\sigma} [(\partial_x\Theta_\sigma)^2+
(\partial_y\Theta_\sigma)^2]. \nonumber
\end{eqnarray}
where $y$ is the imaginary time,  the spin and charge velocities
have been normalised
such that $v_\sigma=v_\rho=1$.  In this convention,
the g-factors have value
$g_\sigma=g_\rho=2$ for a non-interacting system. The case of
quantum wires with $SU(2)$ symmetry corresponds to $g_\sigma=2$.

The electric and magnetic conductances of this
system follow directly from Kubo's formula
\begin{equation}
\label{old}
G_{\rho / \sigma}=\frac{e^2}{h} g_{\rho / \sigma}.
\end{equation}
As discussed in \cite{froh}, this formula might not always 
be  physically relevant,
because of the coupling to the reservoirs. It has
been proposed that $G_\rho=2\frac{e^2}{h}$ might 
always hold, whatever the interactions in the quantum wire.
Of course, this does not mean that the scattering through the 
impurity will be trivial in that case; it will indeed, still be controlled
by the  action written above, the only difference being in the meaning of 
the ``applied voltage''. 
Here, we will simply follow  \cite{kanefish}
for uniformity of notations. Physical results for different reservoirs
configurations can be recovered by rescaling the voltage.

In the presence of a charge impurity
at the origin $x=0$,  the hamiltonian gets an additional piece
\begin{equation}
\delta H=\int dx V(x) (\psi^\dagger_\uparrow \psi_\uparrow+
\psi^\dagger_\downarrow \psi_\downarrow ).
\end{equation}
Here $V(x)$ is a potential which has essentially zero measure outside
$x=0$.  Under the bosonisation rules sketched at the beginning,
this leads to the change of action
\begin{equation}
\delta S=\lambda \int dy \cos\sqrt{\pi}\Theta_\rho
\cos\sqrt{\pi}
\Theta_\sigma ,
\label{chagact}
\end{equation}
where $\lambda\simeq V(2k_F)$.
In the physical case, there is no apparent symmetry that allows us
to restrict to the action (\ref{chagact}) -  the
most general form of the perturbation is
\begin{equation}
\delta S=\int dy \sum_{n_\rho, n_\sigma\atop n_\rho+n_\sigma \ even}
\frac{\lambda_{n_\rho,n_\sigma}}{4} e^{i\sqrt{\pi}(n_\rho
 \Theta_\rho+n_\sigma
\Theta_\sigma)}
\end{equation}
where the $\lambda$'s are real couplings.  The renormalisation
group equations read at  first order
\begin{equation}
\label{rgeq}
\frac{d\lambda_{n_\rho,n_\sigma}}{dl}=
\left( 1- \frac{n_\rho^2}{4} g_\rho-\frac{n_\sigma^2}{4} g_\sigma
\right) \lambda_{n_\rho,n_\sigma }.
\end{equation}
As mentioned before, in this paper we restrict to the case $g_\sigma=2$,
which is the one relevant for physical quantum wires with unbroken $SU(2)$
spin symmetry.
Then, beside the perturbation with $n_\rho=n_\sigma=1$,
the operators $n_\rho=2,n_\sigma=0$ is  relevant for $g_\rho<1$.
The operator with $n_\sigma=2,n_\rho=0$ is always irrelevant.
We will thus further restrict in the following to the (probably 
not physically relevant) case where 
only $\lambda_{11}\neq 0$ or the case where $g_\rho>1$, so we have to take into
account the term $\lambda_{11}$ only, which we simply 
call $\lambda$ in what follows.

In \cite{flsbig}, the equivalent problem for spinless electrons
was exactly solved. The solution required  a folding to transform the
impurity
problem into a boundary problem, and then used recent results
on boundary integrable quantum field theories \cite{GZ} together with
the massless scattering approach. The same folding can easily
be accomplished in the problem with spin. First, it is convenient
to  rescale the fields, writing the action as
\begin{eqnarray}
\label{lateruse}
S&=&S_\rho+S_\sigma+\lambda \int dy \cos\sqrt{\pi g_\rho}\Theta_\rho
\cos\sqrt{2\pi}\Theta_\sigma \\
S_\mu&=&\int dx dy \frac{1}{2} [(\partial_x\Theta_\mu)^2+
(\partial_y \Theta_\mu)^2].
\end{eqnarray}
We then introduce  odd and even fields
\begin{eqnarray}
\Theta^e_\mu&=&\frac{1}{\sqrt{2}} [\Theta_{\mu, L}(x,y )+\Theta_{\mu,
R}
(-x,y)] \\
\Theta^o_\mu&=&\frac{1}{\sqrt{2}} [\Theta_{\mu, L}(x,y )-\Theta_{\mu,
R}
(-x, y )]. \nonumber
\end{eqnarray}
With this, the interaction at $x=0$ involves only the even fields
which have
Neumann boundary conditions.  The odd fields, having Dirichlet
boundary
conditions, completely decouple and do not interact.  Also,
the even field, as defined above, is left moving and we can ``fold" using
\begin{eqnarray}
\theta_{\mu, L}&=& \Theta^e(x+y), \ x<0 \nonumber \\
\theta_{\mu,R}&=& \Theta^e(-x,y), \ x<0.
\end{eqnarray}
We can now express everything in terms of the fields $\theta$ which
is defined on the negative axis, the action becomes
\begin{eqnarray}
\label{bdaction}
S&=&S_0+{\cal B} \nonumber \\
&=&\int_{-\infty}^0 dx \int dy\  \frac{1}{2} \sum_{\mu=\sigma,\rho}
[(\partial_
x\theta_\mu)^2
+(\partial_y\theta_\mu)^2]+\lambda \int dy
\cos\sqrt{\frac{\pi g_\rho}{2}} \theta_\rho (0)
\cos\sqrt{\pi }\theta_\sigma (0).
\label{action}
\end{eqnarray}
The next step in \cite{flsbig} was to use the integrability of the
corresponding boundary quantum field theory - the boundary
sine-Gordon model
 (BSG). In the present case with spin, the
problem, involving two fields, is a priori more complicated.
Indeed it is unlikely that the
problem is  soluble  for general values of $g_\rho, g_\sigma$. However,
in the case $g_\sigma=2$, the problem can
be solved, as we now demonstrate.

The trick to solve a boundary problem such as the one
under consideration here, is to introduce a more general
integrable problem with both bulk and boundary
perturbations. The bulk term is chosen such that the theory remains
integrable thus defining a basis of quasiparticles bulk excitations.
These excitations are interacting through an elastic,
two particles $S$ matrix
derived from the constraints of unitarity, crossing-symmetry and
the Yang-Baxter equation \cite{zamozamo}.
Since the theory with boundary is also integrable, there are
also strong constraints on the boundary reflection matrix $R$
leading to a very simple scattering of these quasiparticles
at the boundary.
When one lets the bulk coupling go to zero, these become
massless quasiparticles which scatter in a simple way at the boundary.
Taking the massless limit is by now a standard procedure and has
been discussed in \cite{muss,warner}.

In our case, we need to find the appropriate bulk perturbation 
for the problem.
The first natural idea
to handle (\ref{bdaction}) is to introduce as a bulk perturbation
\begin{equation}
S_B=\Lambda \int_{-\infty}^0  dx \int dy \ \cos2\sqrt{\pi}\theta_\sigma
\cos\sqrt{2\pi g_\rho}
\theta_\rho
\label{firstpert}
\end{equation}
Note the doubling of the cosine's arguments compared with
(\ref{bdaction}):
this is because, quite generally, an integrable
perturbation of the form $f(\theta^L)f(\theta^R)$
corresponds to an integrable perturbation $f(\theta^R)$ at the
boundary.  Here the
function $f$ is an exponential (to make the action real,
appropriate combinations of exponentials were taken), and, using
Neumann boundary conditions, we wrote $\cos\theta(0)=\cos
2\theta^R(0)$. The problem (\ref{firstpert}) is however not
integrable
for generic values of $g_\rho$ (integrable sub-varieties have been
identified in \cite{lpss,lipatov,fateevss} and will be the subject
of a sequel to this paper), so this approach is not satisfactory.
Fortunately, we can introduce
another bulk perturbation,
\begin{equation}
S_B=\Lambda \int_{-\infty}^0  dx \int dy \
\cos2\sqrt{\pi}\theta^L_\sigma
\cos2\sqrt{\pi}\theta^R_\sigma\cos\sqrt{2\pi g_\rho}
\theta_\rho
\label{secondpert}
\end{equation}
This corresponds to the function $f$ being a cosine function. This
problem  is actually well known to be integrable. To see this, we
can refermionize the $\theta_\sigma$ part. The term
$\cos2\sqrt{\pi}\theta^R_\sigma$
being the sum of a an exponential and its conjugate becomes now a
{\bf Majorana}
fermion, so we have, equivalently,
\begin{equation}
S_B=\Lambda \int_{-\infty}^0  dx \int dy \ \Psi\bar{\Psi}
\cos\sqrt{2\pi g_\rho}
\theta_\rho
\label{goodlook}
\end{equation}
while the free part can be written as a free bosonic hamiltonian,
plus the sum of two decoupled free Majorana fermions, only one of
them appearing into the perturbation
\begin{equation}
S_0=\int_{-\infty}^0 dx \int dy\  \frac{1}{2}
[(\partial_
x\theta_\rho)^2
+(\partial_y\theta_\rho)^2]+
\Psi\bar{\partial}\Psi-\bar{\Psi}\partial
\bar{\Psi}+\chi\bar{\partial}\chi-\bar{\chi}\partial\bar{\chi}.
\end{equation}
Factoring out the
other fermion ($\chi$), we end up with a  standard integrable
$c=3/2$
analog of the sine-Gordon model \cite{BL}. The boundary
perturbation
is also integrable. In terms of fermions, it can be written as
\begin{equation}
\lambda \int dy \
\cos\sqrt{\frac{\pi g_\rho}{2}} \theta_\rho
(0)\left[\Psi(0)+\bar{\Psi}(0)\right]a(y)+{1\over 2}\int dy \left[
\Psi\bar{\Psi}(0)+a\dot{a}\right].
\end{equation}
The last part simply enforces appropriate UV boundary
conditions . The quantum operator $a$ anticommutes with the
fermions, and satisfies $a^2=1$ (see \cite{andre} for a detailed
discussion of bosonization in the presence of a boundary).

In the bulk, the theory can be shown to be integrable by exhibiting
a set of non-local conserved currents
\begin{equation}
J^\pm=\Psi e^{\pm i \theta^R_\rho/\sqrt{2\pi g_\rho}}, \ \bar{J}^\pm=
\overline{\Psi} e^{\pm i \theta^L_\rho/\sqrt{2\pi g_\rho}}
\end{equation}
These currents in turn define  conserved 
charges with  $\widehat{su(2)}_{\hat{q}}$ commutation
relations.  They commute with the charges
constructed from the following local currents
\begin{equation}
G=\Psi \partial \theta_\rho, \ \ \overline{G}=\overline{\Psi}
\ \overline{\partial}\theta_\rho 
\end{equation}
which are also conserved by the perturbation.
Requiring that these symmetries are realized on the asymptotic 
states of the theory  leads to a bulk $S$ matrix
factorized into a a $\widehat{su(2)}_{\hat{q}}$, sine-Gordon
type part describing the scattering of a doublet, $S_{ud}$, 
and a supersymmetric part, which is basically a
kink $S$ matrix with 3 vacua  \cite{nunsmat}.  

It must be mentioned here that there are some subtle  issues
regarding the integrability of the previous field theory - for simplicity
we discuss only the bulk case here, the boundary case being quite similar. 
Although naive power counting following the arguments of \cite{Zamo}
suggests that the theory is integrable at any order, the fact that 
the fermions have integer dimension makes the problem more complicated.
It is generally believed that the integrable theory requires 
an additional term of the form $\Lambda^2\cos 2\sqrt{2\pi g_\rho}\phi$, 
as can be checked 
in the semiclassical case \cite{SM}, or by imposing $N=1$ supersymmetry. 
The universal properties of the model (\ref{goodlook})
are  the same as the one with this additional coupling. Indeed, consider
for instance the computation of the free energy perturbatively. If $a$
is the short distance cut-off, its expansion  involves 
the coupling  $\Lambda (R/a)^{{1\over 2}-
2 g_\rho}$ for every insertion of the first operator, 
and the coupling $\Lambda^2 (R/a)^{1- 8 g_\rho}$ for every insertion
of the second. For $g_\rho$ positive, the integrals are divergent. They could 
either be rendered finite by dimensional regularization (akin to 
taking $g_\rho$ negative), or by using the cut-off. In the latter
case, the divergent part will  only contribute to the  regular, bulk part
(see eg  \cite{CL}). Writing the free energy per unit length
$f=f_{reg} - T \hat{f}$, letting $T=1/R \to 0$
and $\Lambda\to 0$ such that $x\equiv \Lambda R^{1-2g_\rho}$
remains finite, we obtain a function $\hat{f}(x)$ determined fully
by insertions of the first operator, not of the second one.

In fact one can avoid the appeal to this  discussion of the bulk integrable
theory, and $N=1$ supersymmetry, by addressing the problem
from the two channel Kondo model point
of view. It is well established that the anisotropic Kondo problem
is integrable for any number of channels and impurity spin; the 
case at hand corresponds simply to a particular impurity spin - 
a cyclic representation of the quantum algebra $su(2)_q$. 
In particular, for the anisotropic two channel Kondo model, the 
S-matrix of the excitations in the bulk is precisely  the one mentionned
above ($S_{ud} \otimes S_{kink}$). This identification in addition
gives the
impurity S-matrix immediately - this is discussed in more details 
in appendix B.

Recall that the solution of the massless problem proceeds by first
describing  the bulk problem in terms  of massless relativistic excitations
- the limit $\Lambda\to 0$ of the excitations of  (\ref{goodlook}).
The basic massless particles  are then either right or left moving,
with dispersion relation $e=p$ ($e=-p$), and they have a spin quantum
number and a kink quantum number.  We denote them by the doublet
$(u_{a,a\pm 1},d_{a,a\pm 1})$. The label $a$ denotes one of the 3
possible vacua; a kink connects adjacent vacua, and is thus
designated  by a pair of labels (see figure 1).
We use for the  kinematic properties of these basic particles
the same parametrization, $e=m e^\theta$, where $m$ has the dimension
of a mass, although the theory is massless.
\begin{figure}
\vglue 0.4cm\epsfxsize 6cm\centerline{\epsfbox{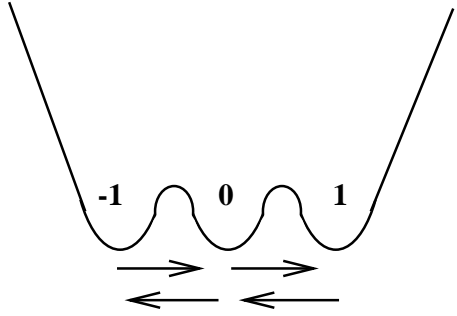}}\vglue 0.4cm
\caption[]{\label{fig1} Structure of the vacuas for the RSOS sector.}
\end{figure}

The particles have factorized scattering, with trivial $LR$
scattering. The LL and
RR scattering are described by the bulk S matrix with  tensor product
structure
\begin{equation}
S=S_{ud}\otimes S_{kink}
\end{equation}
Here $S_{ud}$ is the usual soliton-antisoliton sine-Gordon S-matrix,
corresponding to the coupling ${\beta^2\over 8\pi}\equiv{\gamma\over
\gamma+1}$, ie a quantum group symmetry with
deformation parameter $q=-e^{-i\pi/\gamma}$,
where ${g_\rho\over 4}={\gamma\over 2(\gamma+2)}$, so the total
dimension of the boundary perturbation (using that $g_\sigma=2$)
 is $d={1\over 2}+
{\gamma\over 2(\gamma+2)}={\gamma+1\over\gamma+2}$.
The kink S-matrix
is the well known Restricted Solid on Solid S-matrix for a model with
three vacua.
There may also  be bound states, depending on the value of $\gamma$.
There are no such bound states
if $\gamma$ is larger than one. When $\gamma$ is smaller than one,
there are $n$
of them, where $n$ is the integer part of $1/\gamma$.

The integrability of the model with boundary interaction translates
into the fact that massless particles are reflected  with no particle
production,  in a way described  by a reflection matrix $R$, solution
of the boundary Yang-Baxter equation.
By analogy with the higher spin two-channel Kondo model (see 
appendix B),
we find
\begin{equation}
R=R_{u,d}\otimes 1
\end{equation}
where $R_{u,d}$ is the reflection matrix of the boundary sine-Gordon
problem
at the foregoing coupling $\beta$. It is enough for our purpose to
recall the physical amplitudes
\begin{equation}
\left|R_u^u\right|^2={1\over 1+ e^{-{2\over \gamma}(\theta-\ln
T_B)}}, \
\left|R_u^d\right|^2=1-\left|R_u^u\right|^2
\end{equation}
Here $T_B$ is an energy scale related with the coupling $\lambda$ by
$T_B\propto \lambda ^{-1/(d-1)}\propto \lambda^{\gamma+2}$. Physical
quantities to be discussed below will be expressed in terms of the
squares of R-matrix elements, and therefore expand in powers of
$T_B^{2/\gamma}$. On the other hand,  these quantities expand in
powers of the square of the amplitude
in front of the leading irrelevant operator determining the approach
to the IR fixed point. If the IR fixed point is approached along $\mu O$,
with $O$
of dimension $d$, then $\mu$ has dimension $1-d$. Hence we find
$2(d-1)={2\over \gamma}$ or $d={1\over
\gamma}+1={\gamma+1\over\gamma}={1\over 2}+{\gamma+2\over
2\gamma}={1\over g_\sigma}+{1\over g_\rho}$. Following the
discussion in \cite{kanefish}, this indicates that the IR
fixed point corresponds to ``disconnected leads'', as expected in
that domain of the parameters.

\section{DC transport properties}

The whole logic of \cite{flsbig} can now be implemented. To compute
DC properties, it is enough to treat the massless particles as free
ones
from the point of view of impurity scattering, ie use a Landauer
B\"uttiker type formula.
The massless particles are quantized using TBA equations. The system
of equations, for $\gamma$ an integer, is, setting $\rho_j^h/\rho_j=e^{\epsilon_j}$
\begin{equation}
\epsilon_j(\theta,V/T)= \delta_{j,2} {me^\theta\over T}- K\star\sum_k
N_{jk}\ln\left[1+e^{\mu_k}e^{-\epsilon_k(\theta,V/T)}\right].
\end{equation}
where  the kernel
$K(\theta)\equiv {1\over (2\pi\cosh\theta)}$, and $N_{jk}$
is the
incidence matrix of the following diagram (which is found by
``gluing" the RSOS and the non-diagonal sine-Gordon TBA diagrams)
\begin{figure}
\vglue 0.4cm\epsfxsize 8cm\centerline{\epsfbox{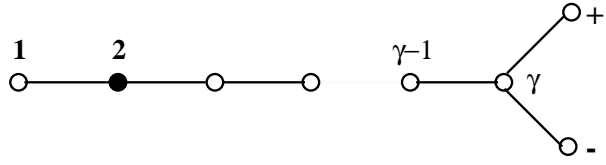}}\vglue 0.4cm
\caption[]{\label{fig2} TBA diagram at zero temperature.}
\end{figure}
%
i.e.\ $N_{jk}=1$ if the nodes $j$ and $k$ are connected, and $0$
otherwise (in particular $N_{jj}=0$). The chemical potential
vanishes, except for the end nodes, where $\mu=\pm (\gamma+1){V\over 2T}$, $V$
the applied voltage.
Here, all the nodes but the one with $j=2$ correspond to
pseudoparticles, which are necessary to diagonalize the S-matrix.

\subsection{Closed form solution at $T=0$}

The simplest case is when $T=0$. Then, a Fermi sea of antisolitons is
formed when a voltage is applied. All the scattering involves these
antisolitons only, so the results can simply be
derived for any value of $\gamma$.  Introduce the quantity $\epsilon$,
solution of
\begin{equation}
\epsilon(\theta)-\int_{-\infty}^A
\Phi(\theta-\theta')\epsilon(\theta')d\theta'={V\over 2}-{m\over
2}e^\theta
\label{epsil}
\end{equation}
Here, $\Phi$ is a kernel with Fourier transform
\begin{equation}
\tilde{\Phi}(\omega)={1\over 4\cosh^2 \pi\omega/2}+{1\over
2\cosh\pi\omega/2}
{\sinh (\gamma-1)\pi\omega/2\over \sinh\gamma\pi\omega/2}
\end{equation}
It is convenient to write it in the form
\begin{equation}
\tilde{\Phi}=1-{\sinh (\gamma+2) \pi\omega/2\over
4\cosh^2\pi\omega/2\sinh\gamma\pi\omega/2}
\end{equation}
At zero temperature, $\epsilon(\theta)$ describes the excitation
energy of the positively charged quasiparticles in the system 
which are the only ones filling the Fermi sea at $V\neq 0, T=0$.
Taking a derivative of (\ref{epsil}) we get the density of these
quasiparticles
\begin{equation}
\rho(\theta)-\int_{-\infty}^A d\theta' \ \rho(\theta') \Phi(\theta-\theta')
=\frac{1}{2h} e^\theta.
\end{equation}
The current then follows from a Boltzmann equation describing
the one by one scattering of these charged excitation on 
the boundary.   There are $\rho(\theta) d\theta$ such particles
in $[\theta, \theta+d\theta]$ and the probability that 
a particle of type $\alpha$ is reflected as one of type $\alpha'$
is $\vert R_\alpha^{\alpha'}(\theta-\log (T_B))\vert^2$.  This leads,
following the arguments of \cite{flsbig} to the current
\begin{equation}
I(V,T_B)=e\int_{-\infty}^A d\theta \ \rho(\theta)
\vert R_u^u(\theta-\log(T_B))\vert^2 .
\label{exact}
\end{equation}
Thus if we manage to determine $A$ and
solve the linear integral equation for $\rho$ we get an exact
evaluation for $I(V,T_B)$.  Since the integral equations are
linear, this is possible by Wiener-Hopf techniques.  

In the previous equations, $A$ is the Fermi rapidity defined by
$\epsilon(A)=0$;
it is the edge of the Fermi sea. $A$ can be determined, and the
foregoing equations
solved, by using a standard Wiener Hopf analysis.
If we write $K(\theta)=\delta(\theta)-\Phi(\theta)$, we have
\begin{equation}
\tilde{K}(\omega)=\frac{1}{N(\omega ) N(-\omega )},
\end{equation}
with
\begin{equation}
N(\omega)=2\pi \sqrt{\frac{\gamma+2}{\gamma}} \frac{
 \Gamma(i(\gamma+2)\omega /2)
e^{i\omega \Lambda}}{\Gamma(i\gamma\omega/2)
\Gamma(1/2+i\omega/2)^2},
\end{equation}
where $\Lambda$ is chosen to preserve the analyticity
in the lower half plane.
Here the conventions for the Fourier transforms are
\begin{equation}
g(\theta )=\int_{-\infty}^\infty \frac{d\omega}{2\pi}
e^{-i \omega \theta} \tilde{g}(\omega), \ \
\tilde{g}(\omega )=\int_{-\infty}^\infty d\theta
e^{i \omega \theta} g(\theta).
\end{equation}
Then following  exactly the same steps as in \cite{flsbig}
we find the solution for the fourier transform
of the density
\begin{equation}
\tilde{\rho}(\omega )=\frac{1}{2 h} \frac{e^{(1+i\omega
)A}}{(1+i\omega)}
N(-i) N(\omega ) .
\end{equation}
The excitation energy of the particles follows
\begin{equation}
\tilde{\epsilon}(\omega) e^{-i\omega A}
=-i \frac{q V}{2} \frac{N(0)N(\omega)}{\omega}
+ \frac{i e^A}{2} \frac{N(-i) N(\omega)}{\omega-i}.
\end{equation}
The condition $\epsilon(A)=0$ is equivalent to
\begin{equation}
\lim_{\omega\rightarrow \infty} \omega \tilde{\epsilon }(i\omega)
 e^{-i\omega A}=0,
\end{equation}
which in turns gives the explicit value of the Fermi rapidity, $A$
\begin{equation}
e^A=e V \frac{N(0)}{N(-i)}.
\end{equation}
We are then left with the evaluation of the current given in
(\ref{exact}).  After a few manipulations we get the exact 
expression
\begin{equation}
I(V,T_B)=\frac{e}{2h} e^A N(-i) \int_{-\infty}^A d\theta
\int_{-\infty}^\infty \frac{d\omega}{2\pi} \frac{e^{-i\omega\theta}}{(1+i\omega)}
\frac{N(\omega)}{1+e^{-2/\gamma(\theta+A-\log T_B)}}.
\end{equation}
This can be expanded in powers of $V/T_B$ to get an IR 
expansion given by
\begin{eqnarray}
I(V,T_B)&=&\frac{ee^A}{2h} N(-i)\sum_{n=1}^\infty (-1)^{n+1}
\frac{N(-2in/\gamma)}{(1+2n/\gamma)} \left(\frac{e^A}{T_B}\right)^{2n/\gamma}
\\
\nonumber 
&=&2\pi\frac{e^2 V}{h} \sum_{n=1}^\infty (-1)^{n+1} 
\frac{\Gamma(2 n/g_\rho)}{(n-1)! \ \Gamma(\frac{1}{2}+\frac{n(2-g_\rho)}{2g_\rho})^2
[1+n(\frac{2-g_\rho}{g_\rho})]}
\left(\frac{e V}{ T_B'} \right)^{n(2-g_\rho)/g_\rho}
\end{eqnarray}
where we have reinstated the dependence on $g_\rho$ in the last
expression and we have introduced the parameter 
\begin{equation}
T_B'=T_B \frac{2\pi}{2-g_\rho}
\end{equation}
for convenience.
We can also separate the integral over rapidities to get an expansion
in the UV.  After a few manipulations we get
\begin{equation}
I(V,T_B)=\frac{e}{2h} N(-i)e^A  PP\int \frac{d\omega}{2\pi}
\frac{N(\omega)}{(1+i\omega)} \left\{
\sum_{n=0}^\infty (-1)^{n+1} \frac{b^n}{2n/\gamma+i\omega}+
\frac{\gamma}{2} \frac{b^{-i\omega \gamma/2}}{
\sin (i \pi \omega \gamma/2)}
\right\}
\end{equation}
with $b=(T_B e^{-A})^{2/g_\rho}$.
We integrate the first sum by closing the contour in the lower
half plane and the second term by closing in the upper half plane.
In the second term, the zeroes of $N(\omega)$ cancels the poles
of $\sinh(\pi \omega \gamma/2)$ and the one at $\omega=i$,
only the poles
$\omega=2ni/(\gamma+2)$ with $n$ integer contribute.
We get
\begin{equation}
I(V,T_B)=\frac{e^2 V}{h} g_\rho+
\sum_{n=1}^\infty  U_{n} \left( \frac{T_B'}{eV}\right)^{n(1-g_\rho/2)},
\label{iis}
\end{equation}
with the coefficients
\begin{equation}
U_{n}=\frac{\pi}{2} \frac{e^2 V}{h} g_\rho^2 \frac{(-1)^n}{(n-1)!}
\frac{\Gamma(ng_\rho/2)}{\Gamma(1/2-n(2-g_\rho)/4)^2
[1-n(2-g_\rho)/2]},
\end{equation}
where again we have reexpressed things in terms of $g_\rho$. From 
these two expressions we can check that there is an exact duality
between the UV expansion and the IR expansion under the 
exchange $g_\rho\rightarrow 4/g_\rho$.  This duality is more transparent
through the relation
\begin{equation}
I(V,T_B',g_\rho)=\frac{e^2 V}{h} g_\rho-\frac{g_\rho^2}{4} I(V,T_B',4/g_\rho).
\end{equation}
In order to make contact with perturbation theory, we need to find the 
relation between the boundary scale $T_B$ and $\lambda$.  This is
done in appendix B using a first order Keldysh computation.

\subsection{Conductance at finite temperature}

To compute the conductance in the case $T\neq 0$ and $\gamma$ an integer
is difficult. The reason is, that the bulk scattering is not diagonal
in the soliton antisoliton basis, and the TBA equations are written, in fact,
for quasiparticles with no definite charge. The transport equation
of \cite{FLS} becomes then ambiguous to use. In the case without spin,
it was possible to conjecture a formula, based on limiting cases, that
reproduced well numerical results, and later was established using
more complicated functional relations. A similar conjecture in the present 
case would be
\begin{equation}
G={\gamma^2\over 2}\int {d\theta\over\cosh^2(\theta-\ln T_B/T)}
\left[{1\over
1+e^{\epsilon_{\gamma+1}(\theta)}}-{1\over 1+e^{\epsilon_{\gamma+1}
(\infty)}}\right]
\label{conjecture}
\end{equation}
It does go to zero in the infrared, as $T/T_B\to 0$. In the 
ultraviolet, when $T/T_B\to\infty$, it is easy to see that $G$  goes to $g_\rho$
as desired. This is because, in that limit, the only contribution to the 
integral
comes from the region $\theta$ near infinity, where the $\epsilon$'s 
go to a constant, in particular, $e^{-\epsilon_{\gamma+1}(-\infty)}\to \gamma+1$ 
(while $e^{-\epsilon_{\gamma+1}(\infty)}\to \gamma-1$).

The case $\gamma=1/n$, $n$ an integer, belongs  to the attractive regime
of the sine-Gordon part of our scattering, and corresponds to the 
case with purely diagonal scattering. It is thus more favorable 
to study the conductance via the approach of \cite{FLS}. However,
while the attractive regime of the ordinary sine-Gordon model
has been much studied, we are not aware of many such studies for the 
$N=1$ sine-Gordon model of interest here. The complete S-matrix was found in 
\cite{breathers}. The TBA has never been written, as far as we know.
 We will present a detailed discussion
of this problem elsewhere. Here, we content ourselves by giving the 
relevant equations
and the corresponding conductance. The diagram looks as follows
\begin{figure}
\vglue 0.4cm\epsfxsize 8cm\centerline{\epsfbox{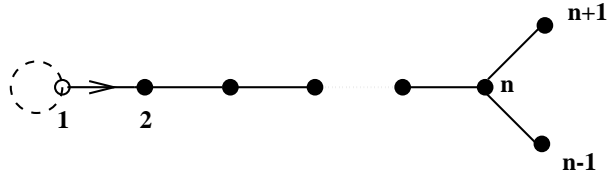}}\vglue 0.4cm
\caption[]{\label{fig3} TBA diagram at finite temperature.}
\end{figure}
%

The TBA equations have the form 
\begin{equation} 
\epsilon_j(\theta,V/T)=K*\sum_k N_{jk}\ln
 \left[1+e^{-\mu_k}e^{\epsilon_k(\theta,
V/T)}\right]
\end{equation}
with $K(\theta)={n\over 2\pi\cosh n\theta}$ and $j=3,\ldots,n+2$, $\mu_{n+1}=
-\mu_{n+2}={V\over 2T}$,
$\mu_k=0$ otherwise. In addition
we have
\begin{equation}
\epsilon_2(\theta,V/T)=K*\left(\ln\left[1+e^{\epsilon_3}\right] 
-\ln\left[1+e^{-\epsilon_1}\right]\right)
\end{equation}
and
\begin{equation}
\epsilon_1=-K*\ln\left[1+e^{\epsilon_2}\right]-K'*\ln\left[1+e^{-\epsilon_1}\right] + \frac{ e^\theta}{T} \frac{\sin\pi/2n}{\cos\pi/2n}.
\end{equation}
Here we have 
\begin{equation}
\tilde{K}={1\over 2\cosh \pi\omega/2n}, \ \ \tilde{K'}={\cosh (n-1)\pi\omega/2n
\over 2\cosh \pi\omega/2n \cosh\pi\omega/2}
\end{equation}
The other mass terms appear in the TBA in the form of asymptotic conditions:
\begin{equation}
\epsilon_j(\theta)\approx \frac{m_j}{T} e^{\theta}
\end{equation}
where  $m_j=2\sin {(j-1)\pi\over 2n}\,\,\,\,j>1$, $m_{n+1}=m_{n+2}=1$. 

We can present, as a quick check of the validity of this TBA,
a computation of the central charge of the theory. Setting
$x\equiv e^{-\epsilon(-\infty)}$ we have the system
\begin{equation}
{1\over x_j}=\prod_k \left(1+{1\over x_k}\right)^{N_{jk}/2}
\end{equation}
supplemented by
\begin{equation}
{1\over x_2}=\left(1+{1\over x_3}\right)^{1/2}(1+x_1)^{-1/2}
\end{equation}
and
\begin{equation}
x_1=\left(1+{1\over x_2}\right)^{1/2}(1+x_1)^{1/2}
\end{equation}
The solution of this system of equations is
\begin{equation}
x_1=3; {1\over x_j}=\left(j-{1\over 2}\right)^2-1, j=2,\ldots,n;{1\over x_{n+1}}=
{1\over x_{n+2}}=n-{1\over 2}
\end{equation}
Similarly, introducing $y\equiv e^{-\epsilon(\infty)}$ we clearly 
have that all the $y$'s vanish, except $y_1=1$. Therefore,
using the general formula
\begin{equation}
c={6\over \pi^2}\sum L\left({x\over 1+x}\right)-L\left({y\over 1+y}\right)
\end{equation}
where $L$ is the dilogarithm function, we find
\begin{equation}
c={6\over\pi^2}\left[ 2L\left({1\over n+{1\over 2}}\right)+\sum_{k=1}^{n-1}
L\left({1\over (k+{1\over 2})^2}\right)+L(3/4)-L(1/2)\right]={3\over 2}
\end{equation}
where we used the identities
\begin{equation}
L\left({1\over x}\right)-L\left({1\over 1+x}\right)={1\over 2} L\left({1\over x^2}\right)
\end{equation}
and 
\begin{equation}
L(x)+L(1-x)={\pi^2\over 6}
\end{equation}
The formula for the linear conductance follows
\begin{equation}
G={n\over 2}\int {d\theta\over \cosh^2 n(\theta-\ln T_B/T)}
{1\over 1+e^{\epsilon_{n+1}(\theta)}}
\end{equation}
with the UV value $G={2\over 2n+1}=g_\rho$. In the limit $T_B/T\to\infty$,
the integral is dominated by the region $\theta\to\infty$ where $\epsilon_n$ 
diverges. It is then legitimate to simply expand the $\cosh$ in the integral, and
one finds that $G$ expands in  powers of $\left({T\over T_B}\right)^{2n}=
\left({T\over T_B}\right)^{{2-g_\rho\over g_\rho}}$, as required.

\section{Conclusions}

To complete the analysis of this problem, we finally compute the boundary 
entropy.
This is easy to do  in the atractive regime.
Since the boundary acts trivially on the kink degrees of freedom,
the boundary scattering of the bound states  is completely given in terms
of the boundary scattering of the sine-Gordon part. This problem has been
studied in \cite{GZ,warner}, and the R-matrices of breathers computed. The 
boundary free energy then follows
\begin{equation}
f_{imp}=-T\sum_{j=2}^{n+2}\int {d\theta\over 2\pi} 
\kappa_j(\theta-\ln(T/T_B))\ln\left(1+e^{-\epsilon_j(\theta)}\right)
\end{equation}
where complete expressions for the $\kappa_j$ can be found
in \cite{warner} (in the latter reference, $\lambda=n+1$). We can thus write
\begin{equation}
s_{UV}-s_{IR}=\sum_{j=2}^n I^{(j)}\ln(1+x_j)+(I^{(+)}+I^{(-)})\ln (1+x_{n+1})
\end{equation}
where $I^{(j)}=\int {d\theta\over 2\pi}\kappa_j(\theta)$. One finds 
thus
\begin{equation}
s_{UV}-s_{IR}
=\sum_{j=2}^n {j-1\over 2}\ln\left[
{(j-1/2)^2\over (j+1/2)(j-3/2)}\right]+{n\over 2}\ln
\left({n+1/2\over n-1/2}\right)={1\over 2}\ln (1+2n).
\end{equation}
We find therefore $s_{UV}-s_{IR}=\ln\sqrt{{4\over g_\rho g_\sigma}}$, 
in agreement with the IR fixed point being made of disconnected leads,
and a computation of the boundary entropies using conformal
partition functions \cite{AL}. 

In the repulsive regime, the computation is more difficult, again
because the scattering on the impurity is non diagonal. However, 
 when $\gamma$ is an integer, we expect, by analogy with the ordinary 
boundary sine-Gordon case,
\begin{equation}
f_{imp}=-T\int {d\theta\over 2\pi}
 {1\over\cosh (\theta-\ln(T_B/T)}\ln \left(1+e^{-\epsilon_{\gamma+1}}\right)
\end{equation}
Hence we find the difference of entropies
\begin{equation}
s_{UV}-s_{IR}={1\over 2}\ln\left(1+e^{-\epsilon_{\gamma+1}(-\infty)}\right)- 
{1\over 2}\ln\left(1+e^{-\epsilon_{\gamma+1}(\infty)}\right)={1\over 2}\ln {\gamma+2\over \gamma}
=\ln\sqrt{{4\over g_\rho g_\sigma}}
\end{equation}
as required. 

The tunneling problem with spin is expected to present many interesting 
features for arbitrary $g_\rho$ and $g_\sigma$. While it presumably 
is not always integrable, we have identified  several integrable varieties
besides the isotropic one just discussed. These include the case $g_\rho+g_\sigma=2$, $g_\rho+g_\sigma=4$, and ${1\over g_\rho}+{1\over g_\sigma}=2$. We hope to  report on the corresponding solutions (and, in some cases, new IR 
fixed points) in a subsequent publication.   

\bigskip

\centerline{\bf Acknowledgements.}

We thank N. Andrei, P. Fendley, T. Hollowood for very useful discussions.  
This work
was supported by the Packard Foundation, the National Young 
investigator program (NSF-PHY-9357207) the DOE
(DE-FG03-84ER40168) and the National Science Foundation
(PHY94-07194).  F. Lesage is also partly supported by 
a canadian NSERC Postdoctoral Fellowship.

\appendix
\section{Analogy with the two channel Kondo problem, and derivation
of the boundary S-matrix}

In this appendix, we discuss along more traditional lines
how the impurity problem with $g_\sigma=2$
can be mapped onto  an anisotropic Kondo-type model.
We start with a general anisotropic {\bf two-channel} Kondo-type
problem with hamiltonian
\begin{equation}
H_0=i \sum_{i,\alpha=1,2}\int_{-\infty}^\infty dx \ \psi^+_{i\alpha}(x)
{\partial\over\partial x}\psi_{i\alpha}(x),
\end{equation}
where we have reformulated the problem so that all quantities are
right movers.  Here $\alpha$ are the spin indices and $i$ indicates
the ``flavor" or fermion type.
The impurity part is
\begin{equation}
H_1=\sum_{\alpha,\beta,i=1}^2\sum_{\lambda=x,y,z} J_\lambda \ S^\lambda
\sigma^\lambda_{\alpha \beta}\psi^+_{i\alpha}(0)\psi_{i\beta}(0)
\end{equation}
where the $\sigma$ are Pauli matrices, while the $S$ are
generators
of the $U_qsl(2)$ algebra
\begin{eqnarray}
\left[S^+,S^-\right]&=&{q^{2S^z}-q^{-2S^z}\over q-q^{-1}}\\
\left[S^z,S^\pm\right]&=&\pm S^\pm .
\end{eqnarray}
The coupling constant are $J_x=J_y\neq J_z$;
the parameter $q$ is related with the anisotropy in a manner to be
described below.

{\bf Abelian} bosonization (as in \cite{EK} which  we follow closely
here) is readily accomplished by introducing
chiral bosons as
\begin{equation}
\psi_{i\alpha}(x)={e^{-i\sqrt{4\pi}\phi^R_{i\alpha}(x)}
\over\sqrt{2\pi a}}.
\end{equation}
The bulk part becomes just a free right moving  boson hamiltonian,
while the impurity part reads
\begin{equation}
H_1={J_z\over\sqrt{\pi}}\sigma^z\left.{\partial\phi^R_s\over \partial
x} \right|_{x=0}
+{J\over \pi a}\left\{
S^+e^{i\sqrt{4\pi}\phi^R_s(0)}+S^-e^{-i\sqrt{4\pi}\phi^R_s(0)}\right\}
\cos\sqrt{4\pi}\phi^R_{sf}(0).
\label{impart}
\end{equation}
In (\ref{impart})  we have expressed the result in terms 
of new boson fields defined by
\begin{eqnarray}
\phi^R_c&=&{1\over 2}\sum_{i,\alpha=1}^2 \phi^R_{i\alpha}\nonumber \\
\phi^R_s&=&{1\over 2}\sum_{i,\alpha=1}^2 
\sigma^z_{\alpha\alpha}\phi^R_{i\alpha} \\
\phi^R_f&=&{1\over 2}\sum_{i,\alpha=1}^2
\sigma^z_{ii}\phi^R_{i\alpha}\nonumber \\
\phi^R_{sf}&=&{1\over 2}\sum_{i,\alpha=1}^2 \sigma^z_{\alpha\alpha}
\sigma^z_{ii}\phi^R_{i\alpha}\nonumber 
\end{eqnarray}
for the charge, spin, flavor and spin-flavor fields respectively.
We find it convenient to absorb the $J_z$ term by performing a
unitary
transformation with $U=\exp [ i{J_z\over \sqrt{\pi}}S^z\phi^R_s(0)]$
that suppresses the $J_z$ term in the impurity  hamiltonian
which reads now
\begin{equation}
H_1=
+{J\over \pi a}\left\{ S^+e^{i\sqrt{\pi g_\rho}\phi^R_s(0)/\sqrt{2}}+
S^-e^{-i\sqrt{\pi g_\rho}\phi^R_s(0)/\sqrt{2}}\right\}
\cos\sqrt{\pi}\phi^R_{sf}(0).
\end{equation}
with  $\sqrt{\pi g_\rho/2}= \sqrt{\pi}-{J_z\over\sqrt{\pi}}$.
Normalizations are
such that the dimension of the impurity operator is
 $d={1\over 2}+{g_\rho\over 4}$. In what follows, we shall use
$g_\rho$
 (instead of $J_z$) as the only dimensionless parameter of the
problem.
The quantum group parameter
for the impurity spin is then $q=e^{i\pi d}$.

The boson fields $\phi_c$ and $\phi_f$ are totally decoupled
from the hamiltonian. If $q$ is a root of unity, and if we chose for
the impurity spin a periodic (cyclic) representation of the quantum
algebra, then,
by following the argument of \cite{FLS}, it is easy to see that the
problem is equivalent to the  ``double''  impurity sine-Gordon model
\begin{equation}
H_1={J'\over \pi a} \cos(\sqrt{4\pi g_\rho}\phi^R_s(0)/\sqrt{2})
\cos \sqrt{4\pi}\phi^R_{sf}(0).
\end{equation}
with $J\propto J'$. Equivalently, we can fold the problem
to recover left and right movers on a half line, transforming
in this way an impurity into a boundary problem
\begin{equation}
H_1={J'\over \pi a} \cos(\sqrt{2\pi g_\rho}\phi_s(0))\cos
2\sqrt{\pi}\phi_{sf}(0).
\end{equation}
 This
coincides with the hamiltonian in the main text, after identification
$\theta_\rho\equiv \phi_s$ and $\theta_\sigma\equiv \phi_{sf}$.

On the other hand, the integrability of the higher spin, multiflavour
Kondo
hamiltonian is well established. Although the isotropic case is
 usually considered,  the proof
immediately generalizes to an impurity spin in an arbitrary quantum
group
representation \cite{FW,AD,W}. To understand the integrability
structure,
it is useful
to perform {\bf non abelian bosonization}. Introduce
the currents
\begin{eqnarray}
{\cal J}&=&\psi_{i\alpha}^+\psi_{i\alpha}\nonumber \\
{\cal J}^\lambda&=&\psi_{i\alpha}^+
\sigma^\lambda_{\alpha\beta}\psi_{i\beta}\\
{\cal J}^a&=&\psi_{i\alpha}^+\sigma^a_{ij}\psi_{i\alpha}\nonumber
\end{eqnarray}
for the $U(1)$ charge, $SU(2)$ spin and $SU(2)$ flavor currents 
respectively.
The bulk hamiltonian is then quadratic in currents
\begin{equation}
H_0={\cal J}{\cal J}+{\cal J}^\lambda{\cal J}^\lambda+{\cal J}^a{\cal
J}^a .
\end{equation}
In the same manner the interaction with the impurity
reads
\begin{equation}
H_1=J({\cal J}^+S^-+{\cal J}^-S^+)+J_z {\cal J}^zS^z.
\end{equation}
Only the spin currents are interacting with the impurity, and we can
forget
 in what follows the free charge boson (a $c=1$ theory) , as well as
the flavor
 current (a $c={3\over 2}$ theory).

The massless scattering description of the general Kondo model
was given in \cite{paul}. The bulk degrees of freedom are the massless
limit of the well known $N=1$ supersymmetric sine-Gordon model. The
effect
of the impurity is described by an R-matrix, whose form
 depends on the impurity spin $s$. In the underscreened case ($s\geq
3/2$),
 it is given by a solution  of the Yang Baxter equation corresponding
to scattering a spin $1/2$ (the $(u,d)$ doublet) through a spin $s-1$
(this renormalisation of the
spin occurs, because the scattering theory is realy an IR
description,
and that, in the IR, $2$ electrons screen the impurity).
The RSOS degrees of freedom scatter then trivially. 
On the other hand, the cyclic representations
behave in many ways like an infinite spin. As a result, the R-matrix
 is  given by an object similar to the underscreened case,
 with the $(u,d)$ doublet
 scattering through a cyclic representation of parameter
 $\hat{q}=-e^{-i\pi/\gamma}$. As discussed in \cite{FLS}, this cyclic
 spin can be gauged away, and one obtains the well known boundary
sine-Gordon
R-matrix for the $(u,d)$ degrees of freedom, as discussed in the
text.

\section{Keldysh Computation.}

In order to compute the differential conductance we need to use
the Keldysh formalism since the system is driven by reservoirs.
To do so, we will use the formulation of the model on the full line
as described by the equations (\ref{lateruse}).  The effect of the
reservoirs can be implemented by shifting the charge fields,
$\Theta_\rho\rightarrow \Theta_\rho+\sqrt{g_\rho/\pi} a(t)/2$ with
$\partial_t a(t)=V$.  Under this prescription the current is
evaluated by taking the functional derivative
\begin{equation}
I(t)=-i \frac{\delta \log Z[a(t)]}{\delta a(t)},
\end{equation}
where we have used conventions in which $\hbar=e=1$.
Using the Keldysh contour, ${\cal C}$, which goes from
$-\infty$ to $\infty$ and then comes back (see figure 3), 
to expand the
partition function we obtain to first order
\begin{figure}
\vglue 0.4cm\epsfxsize 6cm\centerline{\epsfbox{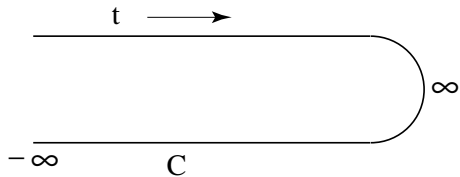}}\vglue 0.4cm
\caption[]{\label{fig4} Contour for the perturbative evaluation.}
\end{figure}
\begin{equation}
I^{(2)}(0)=-i \frac{\lambda^2 g_\rho}{8}\int_{\cal C}dt  \
\sin[\frac{g_\rho}{2}a(t)] \left\{
P_{g_\rho+g_\sigma}^{\mu+}(t)+P_{g_\rho+g_\sigma}^{+\mu}(-t)\right\}
\end{equation}
where $\mu=\pm$ depending on the location of $t$, ie upper part of
the
contour or lower part of the contour.  The functions
$P^{\mu,\mu'}_g(t)$
are the corresponding contraction of the vertex operators time
ordered on the Keldysh contour
\begin{equation}
P^{\pm\pm}_g(t)={1\over (\delta\pm \vert t\vert)^{g/2}}, \ \
P^{\pm \mp}_g(t)={1\over (\delta \mp t)^{g/2}}.
\end{equation}
To this order the result can be found explicitely, it is given by
\begin{equation}
I^{(2)}(0)=-\frac{\pi}{2 \Gamma[(g_\rho+g_\sigma)/2]}
\left(\frac{g_\rho}{2}\right)^{(g_\rho+g_\sigma)/2}
\vert \lambda \vert^2 V^{\frac{g_\rho+g_\sigma}{2}-1} .
\end{equation}
This is in agreement with the Bethe ansatz solution given in the
bulk of the text provided we make the identification
(putting $h=2\pi$ and $e=1$ in the TBA expressions and
$g_\sigma=2$ in the previous perturbative expression)
\begin{equation}
\vert \lambda \vert^2=\frac{8}{g_\rho (2-g_\rho)} \left(
\frac{\Gamma(g_\rho/2+1)}{\Gamma(g_\rho/4)}\right)^2
\left( \frac{2-g_\rho}{\pi g_\rho}\right)^{g_\rho/2}
T_B^{1-g_\rho/2}.
\end{equation}

\end{document}